# Numbers as Data Structures: The Prime Successor Function as Primitive

Ross D. King
Department of Computer Science
Aberystwyth University

## *Abstract*

The symbolic representation of a number should be considered as a data structure, and the choice of data structure depends on the arithmetic operations that are to be performed. Numbers are almost universally represented using position based notations based on exponential powers of a base number – usually 10. This representations is computationally efficient for the standard arithmetic operations, but it is not efficient for factorisation. This has led to a common confusion that factorisation is inherently computationally hard. We propose a new representation of the natural numbers based on bags and using the prime successor function as a primitive – prime bags (PBs). This data structure is more efficient for most arithmetic operations, and enables numbers can be efficiently factored. However, it also has the interesting feature that addition appears to be computationally hard. PBs have an interesting alternative interpretation as partitions of numbers represented in the standard way, and this reveals a novel relationship between prime numbers and the partition function. The PB representation can be extended to rational and irrational numbers, and this provides the most direct proof of the irrationality of the square root of 2. I argue that what needs to be ultimately understood is not the peculiar computation complexity properties of the decimal system (e.g. factorisation), but rather what arithmetical operator trade-offs are generally possible.

## *1. Representing Numbers*

Numbers are abstract objects distinct from their various names and representations. For example the number denoted in decimal notation as "4", is the same abstract object as "100" in binary notation, and "iv" in Roman numerals. Originally numbers had idiosyncratic names (uno, duo, ...; eh'ad, shyim, ...; etc) (Conway & Guy1996). However, as numbers are all somewhat similar, it was found essential to develop systematic ways of naming and symbolically representing them. Historically various cultures have found different more or less elegant solutions to the problem of representing numbers, and it is still instructive to examine these (Knuth, 1981; Gouvea, 2008).

In modern mathematics numbers are almost universally represented as strings, using a position based notation based on exponential powers of the base 10 (Knuth, 1981). In computer science the related binary (base 2) and hexadecimal notations (base 16) are also commonly used. Over time this decimal notation has been extended to represent negative, and rational numbers - where there is commonly used variation based on pairs of decimals – nominator/denominator. Position based systems for representing numbers are now so universal that it is commonly not realised that they are only one choice, among many, of how to represent numbers. This failure to recognise alternatives is especially striking in mathematical/computational problems where the use of a placed based notation is often implicit in the definition of the problem.

One important lesson from computer science is that different representations of the same abstract object may have different computational consequences. For example, a tree may be represented by many different linked structures: binary tree, pre-order sequential, family-order sequential, rings, etc., and "the proper choice of representation depends heavily on what kind of operation we want to perform on the tree" (Knuth, 1968). Although it is generally possible to convert one representation into another, however this conversion may be computationally expensive or intractable.

This paper argues that number representations may be fruitfully considered as data structures, and the choice of data structure depends on the arithmetic operations that we wish to perform.

## 2. The Efficiency of the Place Based Representation of Natural Numbers

The decimal representation is a generally efficient one for standard arithmetic operations (Knuth, 1981; Wiki_complex). The basic properties of the decimal representation are:

- Space efficient: To represent a number of size $n$ it takes space $\Theta(\log_{10} n)$ (all other complexities are time).
- Efficient to test well-formedness: The computational complexity of determining if a $n$-digit number is well formed is time $\Theta(n)$. N.B. it is not trivial to determine whether a number is well formed in some number systems, e.g. Roman numerals.
- Efficient for addition: The computational complexity of the addition of two $n$-digit numbers: using schoolbook addition with carry is $\Theta(n)$. The two most basic operations in arithmetic are addition and multiplication, and the central difficulties of number theory are centred on the way these two operations interact. Addition is almost always considered more important. The reason for this is unclear, for in the physical world multiplicative effects seems equally, if not more, important (Good, 1965).
- Efficient for multiplication: The computational complexity of the multiplication of two $n$-digit numbers using schoolbook long multiplication is $O(n^2)$. The more efficient Karatsuba algorithm has complexity $O(n^{1.585})$.
- Efficient for division: The computational complexity of the division of two $n$-digit numbers using schoolbook long division is $O(n^2)$, and for the Newton's method is $M(n)$ where $M(n)$ stands in for the complexity of the chosen multiplication algorithm.
- Efficient for square roots: The computational complexity of the square roots of an $n$-digit number using Newton's method is $M(n)$.
- Efficient for determining the greatest common factors (GCDs). The computational complexity of determining the greatest common factor of two $n$-digit numbers using the Euclidean algorithm is $O(n^2)$.
- Efficient for deciding whether an integer is Composite or Prime: The computational complexity of determining whether a number is prime or composite is polynominal. The Lenstra and Pomerance deterministic methods is $\tilde{O}((\log n)^6)$ (Lenstra & Pomerance, 2005; Pomerance, 2008). The notation $\tilde{O}(X)$ signifies a bound $c_1 X (\log X)^{c_2}$ for suitable positive constants $c_1, c_2$.
- Inefficient for factorisation: The computational complexity of factoring an $n$-digit number is NP - but not NP complete (Pomerance, 2008; Goldreich & Wigdersom, 2008). The factorisation problem is a famous one, and many cryptographic protocols are based on it. This problem is one where there has often been a lack of precision about specifying the problem: authors have not mentioned an implicit dependency on a placed based representation. For example: "As far as anyone knows, it is a great deal harder to factor a large number $n$ than to compute the greatest common divisor of two large numbers $m$ and $n$" (Knuth, 1981); "no efficient integer factorization algorithm is publicly known" wikipedia (2011). However, there are more careful references: "Here are two simple examples, using decimal notation to code natural numbers: The set of perfect squares is in P, since Newton's method can be used to efficiently approximate square roots. The set of composite numbers is in NP" (Cook, 2011); "... we were interested in positive integers, but formally speaking an algorithm is a function of binary strings., This was not a problem, because there is a convenient an natural way to *encode* integers as binary strings via their usual binary expansion" (Goldreich & Wigdersom, 2008). I profoundly disagree with this quotation as the specific representation used is central to the *problem*.

## 3. A Bag-based Representation of the Natural Numbers

This paper proposes an alternative representation for the numbers based on using bags (multisets), and the prime successor function as primitive. This representation is called "Prime_bags" (PBs). The paper argues that this representation has advantages over the standard decimal representation for some problems.

### 3.1. Bags

A *bag* is an unordered collection of objects (called *elements*) in which elements may occur more than once. The number of times an element occurs in a bag is called its *multiplicity*. A *set* is a bag in which each distinct element has *multiplicity* 1 (Knuth, 1981; Blizard, 1988). A good case can be made that bags are a more realistic physical concept than sets (Blizard, 1988).

Bags may be enumerated by listing their contents between brackets: {}: for example, If B : bag T then B == {a, a, b, c} assigns to B the bag containing the value a twice, the value b once, and the value c once. The number of times x occurs in B is given by the multiplicity function B # x: for example in the bag above, B # a is **2**, i.e. the element "a" occurs twice (the convention in this paper is to use sans serif bold for decimals).

"∈" is the *bag_member* operator. It returns the elements of a bag.

"*" is the *bag_additive_union* operator. If B1, B2: bag T then B1 * B2 is the bag that contains just those elements that occur in either B1 or B2, and the number of times an element x occurs is equal to ((B1 # x) + (B2 # x)): for example, {a} * {b, a} = {b, a, a}. Use of this operator goes back to Weierstrass (Blizard, 1988).

"÷" is the *bag_difference* operator. If B1, B2: bag T then B1 ÷ B2 is the bag that contains for each element the zero-truncated subtraction of the multiplicity functions: for example, {b, a, a} ÷ {a} = {b, a}.

"$^n$" is the *bag_scaling* operator where n is a natural number. If B : bag T then $B^n$ is the bag that contains the elements that occur in B and and the number of times an element x occurs is equal to (n(B # x)): for example, $\{a\}^2$ = {a, a}.

"∩" is the *bag_intersection* operator: if B1, B2: bag T then B1 ∩ B2 is the bag that contains just those elements that occur in both B1 or B2, and the number of times an element x occurs is equal to min((B1 # x), (B2 # x)): for example, {a, a} ∩ {b, a} = {a}.

## 3.2. Prime Bags

Prime bags (PBs) are bags of bags, i.e. bags that contain bags as elements. The PB representation of the natural numbers is based on the following Rules:

Rule 0        {} ∈ PB
The empty bag is a well-formed PB.

Rule 1        (N ∈ PB) → (({{}} * N) ∈ PB)
A well-formed PB is generated by the *bag_additive_union* of the PB that contains only the empty bag to a well-formed PB.

Rule 2        (N ∈ PB, X ∈ N) → ((N ÷ {X}) * {{X}} ∈ PB)
A well-formed PB is generated by replacing an element in a PB with a new element that is the bag containing that element.

Starting with the empty PB (Rule 0) and recursively applying Rules 1 and 2 produces well-formed PBs. The set **PB** is defined as the smallest set containing {} and that is closed under Rule 1 and Rule 2. The structure is a model for the natural numbers ℕ. The interpretation is a follows:
- The first natural number in ℕ is the empty PB {}. This is the number "**1**" in decimal notation.
- The interpretation of Rule 1 is: that adding the empty bag as a member to a PB multiplies it by the decimal number "**2**". Multiplication by **2** is the most basic form of multiplication, it plays an analogous role in multiplication as the addition of the decimal number "**1**" does in addition.
- The interpretation of Rule 2 is that replacing an element of a bag with a bag containing that element is an application of the prime successor function: ps(p) → p + 1 where p is a prime number (the element) and p + 1 is the next prime number in arithmetic succession (the bag containing that element). There are an infinite number of primes, and the prime successor function is primitively recursive.

Starting with the empty-bag, applying Rule 1 to the empty PB produces the PB {{}}. The natural number represented by {{}} (the bag that has as it's only member the empty bag) is the number "**2**" in decimal notation. Rule 2 cannot be applied to the empty bag as it has no members.

Applying Rule 1 to {{}} produces the bag {{},{}}. This is the PB that has two empty bags as members. This represents the number "**4**" in decimal notation. Applying Rule 2 to {{}} produces the PB {{{}}}. This is the the second prime number, the number "**3**" in decimal notation. It is simple to see, thanks to the Fundamental Theorem of Arithmetic, that recursively applying Rules 1 and 2 produces PBs representing all the natural numbers, i.e. it is complete.

Figure 1A illustrates how Rule 1 & 2 can be used to recursively generate the first few natural numbers. This basic bracket based representation is not very human friendly, as it is difficult to distinguish between different numbers of brackets. For this reason it is easier to read the "prime" representation where each member of a PB is the ordered prime, e.g. 1 is the first prime ("**2**" in decimal notation), 2 is the second prime ("**3**" in decimal notation), 3 is the third prime ("**5**" in decimal notation), etc., see Figure 1B). For the remaining of the text we will use the prime representation rather than the basic one. Figure 1C shows the representation of the same numbers as in 1A and 1B in their standard decimal representation.

## *4. The Efficiency of the PB Representation of Natural Numbers*

A PB is an unordered collection of ℕ - as the order of the primes in a PB makes no difference to determining the number. However, when computing with PBs it is convenient to order the bags, i.e. to make them into sequences. It is computational efficient to convert from a unordered bag to an ordered bag: $O(n \log n)$. I will use the convention that decreasing magnitude goes from right to left as in the decimal system.

The basic properties of the ordered PB representation are:

- Space efficient: The representation is efficient, space complexity is $\Theta(\log n)$. An exponential number of natural numbers may represented using a liner number of ordered primes. One way to see this is that infinite set of primes acts as the basis, and the prime number theorem states that the $n$th prime number is approximately equal to $n \ln(n)$. Use of the partition function interpretation provides the exact space complexity (see 7. below).

- Efficient to test well-formedness: The computational complexity of determining if a $n$-digit number is well formed is $\Theta(n)$. Note that this requires the members of the PB to be primes, *not* decimals or other similar position based representations. This is because a PB is only a unique representation of a natural number if its members are primes, and it is not possible to efficiently determine large prime numbers using standard place based representations.

- Efficient for multiplication: The multiplication of two PBs is the bag_additive_union "*" of the two PBs. The computational complexity of the multiplication of two $n$-digit BP numbers is $\Theta(n)$. For example:

    PB                    Decimal
    {1} * {1} = {1, 1}    **2 * 2 = 4**
    {1} * {2} = {2, 1}    **2 * 3 = 6**

The PB representation has many similarities with the the use of logs: multiplication becomes addition, division becomes subtraction, etc., however there is no fixed basis.

- Efficient for division: The division of two PBs is the bag_difference "÷" of the two PBs. The computational complexity of the division of two $n$-digit PB numbers is $\Theta(n)$. For example:

    PB                         Decimal
    {2, 1} ÷ {2} = {1}         **6 ÷ 3 = 2**
    {2, 1} ÷ {1} = {2}         **6 ÷ 2 = 3**
    {2, 1} ÷ {3} is not defined   **6 ÷ 5** has no natural number solution

- Efficient for exponentiation: The exponentiation of a PB is the bag_scaling of the PB. The computational complexity of the exponentiation of a $n$-digit PB number is $\Theta(n)$. For example:

    PB                      Decimal
    $\{1\}^2$ = {1, 1}          $2^2 = 4$           Doubling a PB squares the PB
    $\{1\}^3$ = {1, 1, 1}       $2^3 = 8$           Tripling a PB cubes the PB
    $\{2, 1\}^2$ = {2, 2, 1, 1} $6^2 = 36$
    $\{1, 1\}^{0.5}$ = {1}      $4^{0.5} = 2$       Halving a PB is the square root of the PB

- Efficient for determining the greatest common factors (GCDs). The greatest common factor of two PBs is the bag_intersection "∩" of the two PBs. The computational complexity of the multiplication of two $n$-digit PB numbers is $\Theta(n)$. For example:

    PB                         Decimal

{1, 1} ∩ {2, 1} = {1}.                    The GCD of **4** and **6** is **2**
{3, 1, 1, 1} ∩ {3, 2, 1, 1} = {3, 1, 1}.   The GCD of **40** and **60** is **20**

- Efficient for deciding whether an integer is Composite or Prime: To determine if a number is prime or composite it is only necessary to determine if it has one member or not. The computational complexity of the composition of a *n*-digit BP number is constant.

- Efficient for factorisation: The factors of a PB are simply its members. The computational complexity of the factorisation of a *n*-digit PB number is $\Theta(n)$. N.B. *This is an existence proof that the computational difficulty of factorisation depends on the representation used.*

- Unclear efficiency for addition/subtraction: The obvious disadvantage of the PB representation is that appears computationally complex to determine the arithmetic total ordering of the numbers. This disadvantage, which admittedly is severe, is the reason I believe that PBs and similar representations of numbers have not been studied. I speculate that addition is NP for PBs.

## *5. Addition and the Arithmetic ordering of PBs*

It is computationally easy to determine a partial arithmetic ordering of PBs. If two PBs are not identical, and one is the subset of the other, then the subset PB is obviously smaller. Bertrand's postulate (actually theorem) states that there is always a prime between *n* and 2*n* (Du Sautoy, 2004; Montgomery, & Vaughan, 2007). This means that applying Rule 1 to a PB (multiplying by 2) always produces a number larger that Rule 2 (the prime successor), therefore for a given number of brackets the prime number, the bag with only one member (except the empty bag) is the smallest. The role of Bertrand's postulate in answering a question on the ordering of PBs is one of a number of cases where important questions in number theory translate into the ordering of PBs.

This difference in size when applying Rules 1 & 2 can also be seen by comparing reciprocals (See Figure 1): the series 1/(powers of 2) does converge (Rule 1), while the series 1/primes (Rule 2) does not (Edwards, 1974); Davenport, 2000). It is unclear where in Figure 1 the border between convergence and non-convergence lies.

The difficulty in determining a total ordering comes down to the difficulty in understanding the patterns of primes. This means that there is no simple rule about the ordering of which possible primes to increment will produce a larger number, for example.

{2, 1} **6** → {2, 2} **9**   increment of '1'        {2,1} **6** → {3,1} **10**   increment of '2'
{3,1} **10** → {3,2} **15**   increment of '1'        {3,1} **10** → {4,1} **14**   increment of '3'

Position based systems like that of the decimal and binary systems are based on data structures that construct numbers through the *addition* of powers of a basis (10, 2, etc.). The PB system is based on a data structure that constructs numbers through the *multiplication* of collections of primes. *We hypothesise that there is no efficient conversion between place based representations and PBs.*

## *7. The Partition Function Interpretation*

The systematic generation of PBs using Rules 1 & 2 as shown in Figure 1 produces a well ordering of PBs. This ordering makes explicit an interesting alternative interpretation of PBs as partitions of normal decimal numbers. The Partition function P(n) gives the number of ways of writing the integer as a sum of positive integers, where the order is not considered significant (Kanigel, 1991; Conway and Guy 1996; Wilf, 2000)). By convention, partitions are usually ordered from largest to smallest. For example the partitions of **4** can be written: **4, 3 + 1, 2 +2, 2 + 1 + 1, 1 + 1 + 1 + 1** so P(**4**) = **5**..

Referring to Figure 1 it is clear that the cardinality of the set of PB with a given n pairs of brackets corresponds to the partition function P(n). Each prime number then naturally corresponds to P(n) – 1 composite numbers. The first few values for P(n) are **1, 2, 3, 5, 7, 11, 15, 22, 30, 42**. An asymptotic expression for *P(n)* is given by:

$$P(n) \sim \frac{\exp(\pi\sqrt{2n}/3)}{4n\sqrt{3}} \text{ as } n \to \infty$$

This formula was first obtained by G. H. Hardy and Ramanujan in 1918 (Hardy and Wright 1979). In 1937, Hans Rademacher was able to improve on Hardy and Ramanujan's results by providing a convergent series expression for P(*n*), this series is extremely complex (involving the square root of **2**, π, differential, trigonometric functions, imaginary number) and this belies the apparent simplicity of the function's definition (Wilf, 2000). *The Hardy & Ramanujan formula proves that the PB representation has space complexity Θ(log n).*

## *7. Extensions*

It is simple to extend the PB representation from natural numbers to rational numbers. Just as for place based representations this requires the introduction of a new symbol: "." for decimal fractions, "/" for fractions. To represent rational numbers in PBs we introduce the "-" sign to primes in the bag. This involves changing the definition of bags to allow for each member to have either negative or positive cardinality - bags with negative cardinality seem to have been little studied. It also means that the *bag_difference* (÷) operator to changed to no longer be zero truncated – which is more natural. The interpretation of the negative symbol is as the reciprocal of the prime. For example:

    {1} = **2**                                                    {-1} = **1/2 = 0.5**
    {2} = **3**                                                   {-2} = **1/3 = 0.33..**
    {1, 1} = **2 * 2 = 4**                           {-1, -1} = **1/2 * 1/2 = ¼ = 0.25**

This enables multiplication of two numbers to remain additive_bag_union where, and where primes and their reciprocals may cancel each other out. For example:

    {1} * {-1} = {1, -1} = {}                     **2 * 1/2 = 1**
    {1} * {-2, -2} = {1, -2, -2}               **2 * 1/9 = 2/9**

The division of two rational numbers also remains the bag_difference operator, and division is now the proper inverse of multiplication: {X} ÷ {Y} = {X} * {-Y}. For example:

    ({2, 1} ÷ {1}) = {2} = ({2, 1} * {-1})        **(6 ÷ 2) = 3 = (6 * 1/2)**
    ({2, 1} ÷ {2}) = {1} = ({2, 1} * {-2})        **(6 ÷ 3) = 2 = (6 * 1/3)**
    {2, 1} ÷ {3} = {2, 1, -3)                         **6 ÷ 5 = 6/5**
    {-1} ÷ {-1} = {}                                 **1/2 ÷ 1/2 = 1**

The PB representation of numbers supplies a very simple proof that √2 is irrational. As shown above the sqrt of a PB ({X}$^{0.5}$) is a symmetric binary split of the PB. A simple consideration of symmetry makes it clear that there is no binary split of any rational PB that leaves a remainder {1}. This proof of the irrationality of √2 is related to Lagrange's proof (Square_roots, 2011), but I think the PB representation makes the proof much simpler. It is also trivial to extend the proof to other square roots (only PBs where a binary split is possible), cube roots (only PBs where a tertiary split is possible), etc.

To complete the field of rational numbers it is necessary to introduce symbols for zero and infinity. For infinity we use the standard symbol:

    {∞} = ∞
    {a} * {∞} = {a, ∞} = {∞}          **a * ∞ = ∞**

Zero is more of a problem, as the empty_bag {} represents "1" in decimal and is the identity operator. Therefore for zero we use -∞.

    {-∞} = **1/∞ = 0**
    {a} * {-∞} = {a, -∞} = {-∞}        **a * 0 = 0**

It is also possible to extend the PB representation to deal with irrational numbers. As is the case for placed based representations this requires the introduction of a new symbol (e.g. "√" for roots). To represent rational numbers in PBs we introduce the "/" sign to primes in the bag. For example:

    {1/2} = √**2**,                      This is the PB that when added to itself is {1}
    {1/3} = ∛**2**,
    {1/2, 1/2} = √**2** * √**2 = 2 = {1}**     Prime fractions in the bag with the same denominator may be added.

$\{1/3, 1/3, 1/3\} = \sqrt[3]{2} * \sqrt[3]{2} * \sqrt[3]{2} = \mathbf{2} = \{1\}$
$\{-1/2\} = \mathbf{1/\sqrt{2}}$,
$\{-1/3\} = \mathbf{1/\sqrt[3]{3}}$,

It is also simple to extend PBs to represent negative numbers. This requires the introduction a negative symbol for the bag. This might be considered to overload the symbol "-" but it seems the most natural usage. For example:
- $-\{\} = \mathbf{-1}$
- $-\{1\} = \mathbf{-2}$
- $-\{-1\} = \mathbf{-1/2}$
- $-\{1,1\} = -(\mathbf{2 * 2}) = \mathbf{-4}$
- $-\{-1, -1\} = -(\mathbf{1/2 * 1/2}) = \mathbf{-1/4}$

The additive bag_union operator needs to be adapted to ensure the normal multiplication rules for negative numbers:
- $-\{1\} * \{1\} = -\{1, 1\}$
- $-\{1\} * -\{1\} = \{1, 1\}$

A similar approach can be used to extend PBs to represent imaginary numbers. For example:
- $i\{\} = \mathbf{i}$
- $i\{1\} = \mathbf{2i}$
- $i\{-1\} = \mathbf{1/2i}$
- $i\{1,1\} = \mathbf{4i}$

Again the additive bag_union operator needs to be adapted to ensure the normal multiplication rules for imaginary numbers:
- $i\{1\} * \{1\} = i\{1, 1\}$
- $i\{1\} * -\{1\} = -i\{1, 1\}$
- $i\{1\} * i\{1\} = -\{1, 1\}$

Finally it also possible to represent transcendental numbers. Using the Euler product formula with $n = \mathbf{2}$ produces a pretty definition of π (Edwards, 1974):
$\pi^2 = \{1, 2\} * (\{\} \div ((\{\} -\{-1\}^2) * (\{\} -\{-2\}^2) * (\{\} -\{-3\}^2) \dots ))$

## 8. Discussion

### 8.1. Unique and Non-Unique Representations

Most number systems are based on a unique symbolic representation for each number. However this is not essential, e.g. IIII and IV represent the same number in Roman numerals, and 5.0 and 4.999.. are the same number in the decimal system. It is a truism in computer science it is generally possible to trade space for time. This suggests that it could be possible to form number systems where the mapping from abstract number to symbolic representation is 1 to n (i.e. use a data structure that is less efficient in space), and which are faster to compute arithmetic operations (i.e. a data structure that is more efficient in time) (Anderson, 1971).

Two related bag based number representations shed light on the advantages/disadvantages of the decimal and PB systems, and the advantages/disadvantages of using unique and non-unique representations. The first is an *addition* based system where the elements 0, 1, 2, ... represent powers of 10. For example: $\{0\}$ = **1**, $\{0, 0\} = \mathbf{2}$, $\{0, 0, 0\} = \mathbf{3}$, $\{1\} = \mathbf{10}$, $\{1, 1\} = \mathbf{20}$, $\{1, 0\} = \mathbf{11}$, etc. If the bags are sorted (which can be done efficiently) this systems resembles the decimal system, but it differs in that there is not a unique representation for every abstract number, e.g. both $\{0, 0, 0, 0, 0, 0, 0, 0, 0, 0\}$ and $\{1\}$ represent **10**. This bag based system resembles coins in a pocket: if you have ten individual penny coins in your pocket they do not miraculously convert themselves into one 10 pence coin. The non-uniqueness of the mapping of the quantity of money to pockets of coins does not present any problems to commerce, and I do not think it presents any problem to arithmetic. In this system addition and subtraction are efficient: addition is

bag_additive_union and subtraction is bag_difference. Multiplication is also much simpler than schoolbook multiplication e.g.

**2** x **11** = {0, 0} x {1, 0} = ({0, 0} x {1}) + ({0, 0} x {0}) = {1, 1} + {0, 0} = {1, 1, 0, 0} = **22**

It is also efficient to convert from this bag based representation to a standard place based system.

The other related bag based number representations is a multiplication based system. It differs from the PB system in that the objects in the bag are normal integers rather than primes. For example {2, 2, 2} = **8**, {4, 2} = **8**, {9, 9} = **81**, etc. Unlike the PB representation there is not a unique representation for every abstract number. In this representation many of the efficiency results for PBs carry over, i.e. for multiplication, division, exponention. It is also efficient to convert from this bag based representation to a standard place based system, and thus compute additions and subtractions.

## 8.2. The Relationship Between Number Representation and Arithmetic Efficiency

This paper does not propose to use bags as an alternative formal definition of numbers, although this is possible (Blizard, 1988), the standard formal definition of a bag assumes the natural numbers: as a 2-tuple($A$, $m$) where $A$ is some set and $m$: $A \rightarrow \mathbb{N}$ is a function from $A$ to the set $\mathbb{N}$. The motivation of the paper is rather to better understand the relationship between the representation of a number and what can be efficiently computed. I argue that the choice of representation for a number is a choice of data structure that makes some computations easy, and perhaps others difficult. The standard position based representation (decimal, binary, etc.) is a good engineered solution that make the basic arithmetic operations efficient (e.g. addition, multiplication), but which also seem to prevent the efficient computation of others (e.g. factorisation). The PB representation was created to demonstrate that other representations are possible that optimised for other computations (e.g. multiplication, factorisation, etc.), but are inefficient at others (e.g. addition and subtraction).

In this regard it is perhaps instructive that the only polynomial time algorithm known for factorisation (the Shor algorithm) is based on using a quantum computer (Nielson & Chuang, 2001). The true ontological interpretation of quantum mechanics is unclear, however the superposition of states in quantum computers may be interpreted as enabling the use of data structures not possible in classical computers; and I argue that is is this that enables the speed up in factorisation - assuming quantum mechanics is correct, as no quantum computer has yet been built.

To conclude there is a common incorrect assumption that the only way to represent numbers is using place based systems, and this has led to imprecision in describing the computational complexity of arithmetic problems; for example, the problem of efficiently factorisation large integers is not a fundamentally hard problem. *What needs to be ultimately understood is not the peculiar computation complexity properties of the decimal (or PB) system, but rather what arithmetic operator trade-offs are possible with different data structures*.

**Figure 1**

         {}

        {{}}

    {{{}}}     {{}, {}}

  {{{{}}}}    {{{}}, {}}    {{}, {}, {}}

{{{{{}}}}} {{{{}}, {}} {{{}}, {{}}}  {{{}}, {}, {}}   {{}, {}, {}, {}}

A) The basic representation. Each level shows the application of Rule 1 & 2 to the previous level.

         {}

        {1}

     {2}     {1, 1}

   {3}    {2, 1}    {1, 1, 1}

 {4}   {3, 1} {2, 2}  {2, 1, 1}    {1, 1, 1, 1}

B) The prime representation.

         1

        2

     3     4

   5     6     8

7   10  9   12     16

C) The decimal representation.